\documentclass[lettersize,journal]{IEEEtran}

\usepackage{hyperref}
\usepackage{amsmath, colortbl}
\usepackage{float}
\usepackage{cleveref}
\crefname{equation}{Eq.}{Eqs.}
\crefname{figure}{Fig.}{Figs.}
\crefname{table}{Tab.}{Tabs.}
\usepackage{enumitem}

\usepackage{eurosym}                

\usepackage{amsmath,amsfonts}
\usepackage{algorithmic}
\usepackage{algorithm}
\usepackage{array}
\usepackage[caption=false,font=normalsize,labelfont=sf,textfont=sf]{subfig}
\usepackage{textcomp}
\usepackage{stfloats}
\usepackage{url}
\usepackage{verbatim}
\usepackage{graphicx}
\usepackage{cite}
\begin{document}

\title{A day-ahead market model for power systems: benchmarking and security implications}

\author{Andrej~Stankovski, Blazhe Gjorgiev, James Ciyu Qin, and Giovanni Sansavini, Member, IEEE
\thanks{A. Stankovski, B. Gjorgiev, J. C. Qin, and G. Sansavini are with the Reliability and Risk Engineering Laboratory, Institute of Energy and Process Engineering, ETH Zurich, 8092 Zurich, Switzerland.}
\thanks{Corresponding author: G. Sansavini (sansavig@ethz.ch).}}


\IEEEpubid{}

\maketitle

\begin{abstract}
Power system security assessments, e.g. via cascading outage models, often use operational set-points based on optimal power flow (OPF) dispatch. However, driven by cost minimization, OPF provides an ideal, albeit unrealistic, clearing of the generating units that disregards the complex interactions among market participants. In addition, existing market modeling tools often utilize economic dispatch and unit commitment to minimize total system costs, often disregarding the profit-driven behavior of market participants. The security of the system, therefore, may be overestimated. To address this gap, we introduce a social-welfare-based day-ahead market-clearing model. The security implications are analyzed using Cascades, a model for cascading failure analysis. We apply this model to the IEEE-118 bus system with three independent control zones. The results show that market dispatch leads to an increase in demand not served (DNS) of up to 80\% higher than OPF, highlighting a significant security overestimation. This is especially pronounced in large-scale cascading events with DNS above 100MW. A key driver is the increased dispatch of storage and gas units, which can place the system in critical operating conditions. Operators can use this information to properly estimate the impact of the market on system security and plan efficient expansion strategies.
\end{abstract}


\section{Introduction}
\IEEEPARstart{T}{he} increased adoption of intermittent renewable power sources, the phase-out of dispatchable power generation technologies, and increased demand due to electrification have exposed the system to unprecedented challenges \cite{Heuberger2018Real-WorldSystems}. As a result, the system frequently operates near critical loading conditions~\cite{Alhelou2019AChallenges}, making the safe operation of the system more challenging. Power market dynamics have also shown their potential to trigger grid instabilities, but their impact is often overlooked. In 2019, German Transmission System Operators (TSOs) were forced to resolve a system imbalance of nearly 6,000 MW that was caused by a sharp increase in intraday prices following an inaccurate renewable forecast  \cite{50Hertz2019Investigation2019}. To resolve the imbalance, the TSOs fully activated the available reserves, shed demand, and procured balancing energy from neighboring countries. Only six days later, a similar incident occurred, causing an imbalance of nearly 10,000 MW. In the same year, a network capacity forecasting error caused 1/3 of Switzerland's generation to be exported to the German market, resulting in numerous overloads and threatening the security of the system~\cite{Swissgrid2019Network2019}. These events are among the numerous near misses, underscoring the importance of understanding the impact of market clearing on system security, despite ever-increasing forecast accuracy and refined market design. To this end, many security models assume OPF-based dispatch. Driven by cost minimization and informed by the grid capacity, the OPF provides an ideal, albeit unrealistic, dispatch of the generating units, disregarding the complex interactions between market participants. In reality, the market is completely decoupled, and the participants are unaware of the intra-zonal grid constraints. This can result in uneven loading across the grid, increasing the risk of line congestion and overloads. Therefore, the security of the system can be overestimated when relying on OPF. In this manuscript, we aim to 1) create a realistic representation of the market mechanisms; 2) assess the market impact on system security; 3) quantify how much OPF overestimates system security.

\subsection{Market models in literature}
\label{euroEm_lit_market}
The objective of the market is to maximize social welfare, outlined as the surplus of the consumers minus the surplus of the producers. Although current electricity markets are quite competitive~\cite{Graf2013MeasuringElectricity}, inefficiencies can occur in hours with very high demand and low supply. Anticipating this, market participants can place strategic bids to maximize their profits, resulting in a market clearing that deviates from the cost-optimal solution. Capturing this behavior is difficult, however, and existing market models tend to employ methods that imply perfect competition between market participants (e.g. economic dispatch, unit commitment). 
For instance, Meibom et al. ~\cite{Meibom2006WilmarDocumentation} propose a unit commitment optimization problem to emulate the decisions of the market participants. EMMA, a market modeling framework presented in~\cite{Hirth2018WhatAnalysis}, uses economic dispatch to analyze market behavior and develop optimal investment strategies. Enerpol, an integrated electricity and gas model~\cite{Singh2015ImprovedSystems}, analyzes the behavior of market participants and uses AC optimal power flow (OPF) to simulate market clearing. PyPSA, an energy system modeling framework, also provides market clearing using an economic dispatch ~\cite{Horsch2018PyPSA-Eur:System}. An article examining the performance of five market models for a coal phase-out analysis is presented in~\cite{Postges2021PhasingModels}. The outlined methods utilize economic dispatch or unit commitment.

Attempting to capture this behavior,~\cite{Ruiz2009PoolPrices} propose the use of bi-level optimization problems to represent the strategic bidding as a leader-follower problem, where the unit (leader) optimizes the volume or cost of their bids while the market (follower) reacts to the changing bids. \cite{Khazali2021Risk-AwareTheory} expands on this idea, developing bi-level optimization models that focus on profit maximization of storage and renewable energy sources. These models are especially valuable for strategic placement of storage arbitrage, as highlighted in~\cite{Khaloie2024Day-AheadMarkets} and~\cite{Khaloie2025HybridModel}. However, the models assume perfect foresight of the market outcome and employ a unit commitment formulation, forcing the market to respect the ramp and operating limits of the units. In contrast, the clearing of liberal power markets (e.g., EPEX SPOT) is solely driven by the merit order. Respecting the unit constraints is thus the responsibility of the asset owner. In addition, solving bi-level optimization problems for large-scale applications is computationally expensive and often intractable.

\subsection{Power system security}
Cascading failures are the most significant threat to power system security ~\cite{Stankovski2023PowerEvidence}. To investigate the cascading failure mechanism in detail and assess power system security, cascading failure models have been developed and are now seen as the standard for in-depth security assessment. Flow-based models are among the most popular models of this type, utilizing DC or AC-based power flow to identify congestions in the grid. One of the first models of this type is published by Dobson et al.~\cite{Dobson2001AnBlackouts}. The Manchester model~\cite{Nedic2006CriticalityModel} introduced AC optimal power flow assessments, methods that are computationally costly, but more accurate as they take reactive power and losses into account. Additionally, this model expands the algorithm by introducing random outages. The works presented in~\cite{David2020QuantitativeSystems},~\cite{Henneaux2018BenchmarkingMethodologies}, and~\cite{Bialek2016BenchmarkingTools} benchmark and assess the performance of several popular cascading failure models from the research and industry fields, highlighting their strengths and weaknesses. Informed by the extensive literature in this field, Cascades, a large-scale model with multi-zonal frequency regulation capabilities, was developed~\cite{Gjorgiev2021Cascade-risk-informedSystems, Stankovski2022Multi-zonalSystems}. Although efforts have been made in recent years to apply market dispatch to security models~\cite{Gjorgiev2022Nexus-e:Systems, Garrison2018CombiningSwitzerland}, these models are limited to smaller systems and utilize economic dispatch. To the best of the author's knowledge, a model that employs unbiased day-ahead clearing to large interconnected systems is lacking.

\subsection{Contribution of this work}
The literature outlines the following research gaps: 1) a market dispatch model that simulates the profit-driven behavior of participants while maintaining a social welfare-based solution is lacking, and 2) the impact of different dispatch models on system security is not understood. 
We address these gaps by introducing a day-ahead electricity market dispatch model (DAM) inspired by the behavior of the European electricity market (EPEX SPOT~\cite{EPEXSPOT2021FundamentalsMarkets}). We benchmark the performance of the model against an economic dispatch (ED) and a unit commitment model (UC), commonly used in literature. The day-ahead model is part of a common \textbf{ Euro}pean \textbf{E}lectricity \textbf{M}arket modeling framework, referred to as EuroEM. The DAM model is a "copper-plate" dispatch model, disregarding the power grid physics; therefore, we utilize a redispatch model presented in~\cite{Qin2025PreventiveFramework} to ensure the grid constraints are satisfied by mimicking the actions of the TSO. In addition, the model allows for assessing congestion management costs, a highly contentious point in modern power markets~\cite{ThemaConsultingGroupRedispatchReport}. The redispatch model serves as an interface between the different generation dispatch models and the system security analysis tool, Cascades.
We use the EuroEM framework to answer the following research questions: i) Can the DAM capture the competitive behavior of the real power market? ii) How does the DAM compare to the ED and UC, the current literature standards? iii) What is the impact of market dispatch on system security? iv) Is the security overestimated in OPF-based analyses? By answering these research questions, we contribute in two ways: 1) we present a market modeling framework capable of capturing the competitive behavior of the real electricity market, and 2) we perform a market-informed cascading failure analysis, highlighting deficiencies in the traditional OPF-based analysis.

We perform the market and system security analyses on a modified IEEE-118 bus system with three independent control zones. Our results show that all dispatch models are significantly worse for system security compared to an OPF clearing. In addition, DAM degrades system security more than ED and UC. The findings suggest that greater utilization of storage and gas units can lead to a less favorable distribution of generation, placing the system in more critical conditions. Therefore, our model can help identify critical scenarios that are often overlooked in traditional OPF-based security simulations. Operators can use this information to properly allocate reserves and perform efficient expansion planning strategies, preparing the system for changes in the generation mix and market regulations.

The rest of the paper is structured as follows: Section~\ref{sec:methodology} describes the methodology; Section~\ref{chapter6_case_study} defines the case study and the test system utilized in this work; Section~\ref{sec:results} outlines the main results of the study; and finally, Section~\ref{sec:conclusions} provides conclusions and outlook for our future work.

\section{Method}
\label{sec:methodology}
\begin{figure}[!t]
    \centering
    \includegraphics[width=0.5\textwidth]{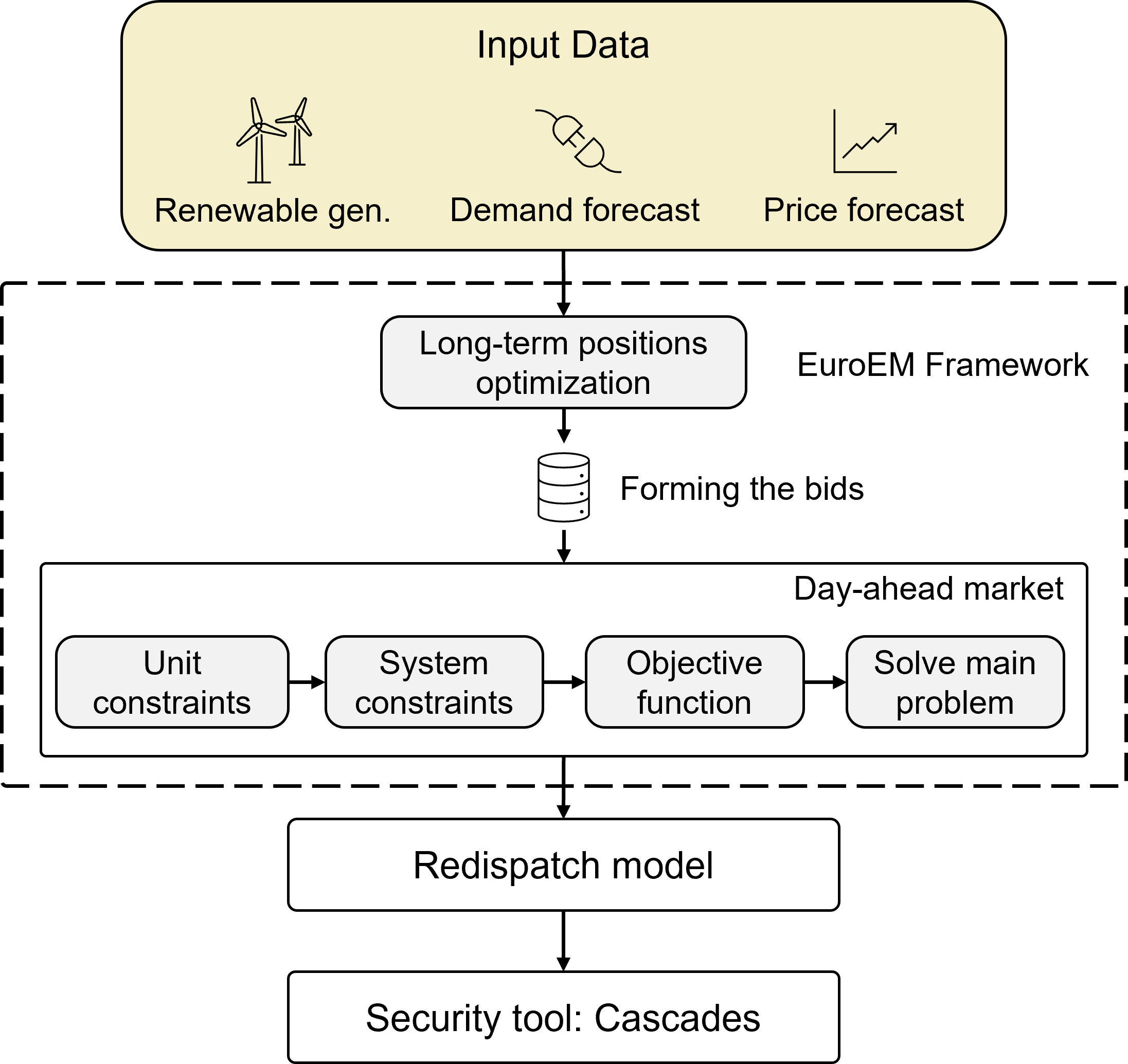}
    \vspace{-0.5cm}
    \caption{Overview of the EuroEM algorithm. The long-term positions optimization sets the bidding structure for the day-ahead market model (DAM). The results from the DAM clearing are passed to the redispatch and security models.}
    \label{fig:EuroEM_algorithm}
\end{figure}

Figure~\ref{fig:EuroEM_algorithm} shows an overview of the 3-step algorithm, which includes the day-ahead market (DAM) clearing of the (EuroEM) framework, the redispatch model, and the cascades model. For DAM clearing, a prerequisite long-term positions optimization (LTPO) step is performed to set the bidding strategy of the participants. Afterwards, we define the corresponding constraints and the objective function and solve the problem. The solution includes the clearing price, cross-border trade, and positions of the generating units. The analysis is performed on a control-zone level, where only the inter-zonal transfer capacity is of interest. This is inspired by the European electricity markets (i.e., EPEX SPOT~\cite{EPEXSPOT2021FundamentalsMarkets}), where the market participants are not aware of the intra-zonal grid constraints. The market dispatch is then redispatched and used by the Cascades model for security assessment.

Cascades is a quasi-steady state cascading failure analysis model. This model can perform both single-zone~\cite{Gjorgiev2022IdentifyingAnalysis} and multi-zonal~\cite{Stankovski2022Multi-zonalSystems} security assessments, as well as devise optimal expansion planning strategies~\cite{Gjorgiev2021Cascade-risk-informedSystems, Gjorgiev2022Nexus-e:Systems}. Similar to other cascading failure models, the standard Cascades algorithm depends on AC/DC OPF to set the initial conditions of the study. The model has been validated and calibrated using historical data from the Western Electricity Coordinating Council (WECC) power grid~\cite{Gjorgiev2019CalibrationAssessment}.

In Section~\ref{sec:methodology}, we focus on the market dispatch model and its constituent modules. The economic dispatch and unit commitment models are commonly used in the literature, and their mathematical formulations are presented in the Supplementary material. The daily redispatch is performed via~\cite{Qin2025PreventiveFramework}.

\subsection{Long-term positions optimization module}
\label{EuroEM_dam_ltpo}

The long-term positions optimization (LTPO) module performs an hourly bid stacking for the units based on the expected market outcome. The input data for this module is the technical specifications of thermal and storage units and the price forecast. The time horizon of the analysis is presented in hourly resolution, denoted by $T$, while $\rho_t$ is the forecasted price. The price forecast varies for each participant based on their bidding zone and the level of forecast uncertainty.

The goal of the optimization is to represent the profit-driven nature of the market participants. We employ a mixed-integer linear programming (MILP) model with a profit maximization objective function, subject to technical constraints (Eq. \ref{EuroEM_eq_ltpo}). The objective is to maximize the profit of each market participant based on the price forecast, costs and their technical limitations. The thermal constraints are presented in Eqs. \ref{EuroEM_eq_ltpo}b - \cref{EuroEM_eq_ltpo}h. $P_{g,t}$ represents the hourly generation of the thermal unit $g$, while $N_G$ denotes the set of thermal units in the system. The hourly power discharge and charge of storage units are represented with $PD_{s,t}$ and $PC_{s,t}$, respectively. The VOM costs of storage units are presented with $b_s$ in the objective function (Eq. \ref{EuroEM_eq_ltpo}a). Eqs. \ref{EuroEM_eq_ltpo}b - \ref{EuroEM_eq_ltpo}d are the operating and ramp constraints, where $RU_g$ and $RD_g$ represent the ramp up and down limits, respectively. The minimum and maximum operating limits of the units are presented with $P_{g}^{min}$ and $P_{g}^{max}$. The marginal, start-up, and shutdown costs of thermal units are represented by $ b_g$, $ b^{on}_{g}$, and $b^{off}_{g}$, respectively. The marginal costs include variable operations and maintenance (VOM) costs, dynamic fuel costs, and $CO_2$ emission costs. The operating status of the thermal units is presented with the binary variable $\nu^{stat}_{g,t}$, while $\nu^{on}_{g,t}$ and $\nu^{off}_{g,t}$ are the start-up and shut-down binary decision variables, respectively. The unit status is updated by Eq. \ref{EuroEM_eq_ltpo}e, while Eq. \ref{EuroEM_eq_ltpo}f prevents simultaneous start-up and shut-down signals.

\begin{subequations}
\begin{align}
& \max_{P_g, PD_s, PC_s} \sum_{t \in T} \sum_{g \in N_G}(P_{g,t}\cdot \rho_{g, t} - P_{g,t} \cdot b_g - \nu^{on}_{g,t} \cdot b^{on}_{g} \nonumber \\  
& - \nu^{off}_{g,t}\cdot b^{off}_{g}) + \sum_{t \in T} \sum_{s \in N_S}((PD_{s,t} - PC_{s,t})\cdot \rho_{s,t} \nonumber \\  
& - (PD_{s,t} + PC_{s,t}) \cdot b_s) \label{objective} \\
& \text{s.t.} \nonumber  \\
& Thermal \: constraints \nonumber  \\
& P_{g}^{min} \cdot \nu^{stat}_{g} \leq P_{g, t} \leq P_{g}^{max} \cdot \nu^{stat}_{g}, \quad \forall \{g, t\} \\
& P_{g, t+1} - P_{g, t} \leq RU_{g}, \quad  \forall \{g, t\}\\
& P_{g, t} - P_{g, t+1} \leq RD_{g}, \quad  \forall \{g, t\}\\
& \nu^{stat}_{g,t+1} - \nu^{stat}_{g,t} = \nu^{on}_{g,t} - \nu^{off}_{g,t}, \quad  \forall \{g, t\}\\
& \nu^{on}_{g,t} + \nu^{off}_{g,t} \leq 1 \quad  \forall \{g, t\}\\
& Storage \: constraints \nonumber  \\
& PD_{s}^{min} \leq PD_{s, t} \leq PD_{s}^{max}, \quad \forall \{s, t\}\\
& PC_{s}^{min} \leq PC_{s, t} \leq PC_{s}^{max}, \quad \forall \{s, t\}\\
& E_{s}^{min} \leq E_{s, t} \leq E_{s}^{max}, \quad \forall \{s, t\}\\
& E_{s, t} = (1 - SD_{rate}) \cdot E_{s, t-1} + \eta_{C, s} \cdot PC_{s,t} \nonumber \\
& \hspace{2.0em} - 1/\eta_{D, s} \cdot PD_{s,t} + SI_{s, t} - SW_{s, t}, \quad \forall \{s, t\} \\
& 0 \leq SW_{s, t} \leq SW_{s}^{max}, \quad \forall \{s, t\}\\
& E_{s, t = T} \geq E_{s, t=0}, \quad \forall s
\end{align}
\label{EuroEM_eq_ltpo}
\end{subequations}

The storage unit constraints are presented with Eqs. \ref{EuroEM_eq_ltpo}g - \ref{EuroEM_eq_ltpo}l, where Eqs. \ref{EuroEM_eq_ltpo}g - \ref{EuroEM_eq_ltpo}h describe the technical operating limits.The minimum and maximum discharge/charge limits are presented with $PD_{s}^{min}/PC_{s}^{min}$ and $PD_{s}^{max}/PC_{s}^{max}$, respectively. Eq. \ref{EuroEM_eq_ltpo}i shows the storage level constraints, mandating the level to be maintained between the permissible $E^{min}_s$ and $E^{max}_s$ values. The hourly storage level (state of charge) of the units is presented with $E_{s,t}$, while $N_S $ is the set of storage units in the system. The state of charge constraint is presented in \ref{EuroEM_eq_ltpo}j. The difference between the charging ($\eta_{C, s}$) and discharging ($\eta_{D, s}$) efficiencies discourages simultaneous charging and discharging. We employ this approach to avoid the computational expense of additional binary variables. Due to the fast-operating nature of storage units, we exclude ramp constraints. The lowest temporal resolution in the EuroEM framework is one hour, which is sufficient for units of this type. $SD_{rate}$, $SI_{s,t}$, and $SW_{s,t}$ in Eq.\ref{EuroEM_eq_ltpo}j are the self-discharge rate, inflows, and waste spillage, respectively. Table~\ref{EuroEM_table_storage} specifies the inputs for all storage units based on their operating technology. Lastly, Eq.~\ref{EuroEM_eq_ltpo}l mandates the storage level of each unit at the end of the time horizon to be at least equal to the level at the beginning of the optimization. This constraint motivates the cycling of units with charging capabilities, as well as ensuring a predictable water level at the beginning of the year for large hydro dam units.

\begin{table*}
\begin{center}
\centering    
\caption{Relevant inputs for storage units based on the operating technology. Hydro pumped units with daily operation refer to pumped units with no natural inflows.}
\label{EuroEM_table_storage}
\begin{tabular}{| c | c | c | c |}
\hline
\textbf{Technology }&\textbf{Non-zero inputs} & \textbf{Zero inputs} \\
\hline
Hydro dam & Inflows $SI$, waste outflows $SW$ & Charging limits $PC$, self-discharge $SD$\\
Hydro pumped storage & Charging limits $PC$, inflows $SI$, waste outflows $SW$ & Self-discharge $SD$\\
Hydro pumped, daily operation & Charging limits $PC$, waste outflows $SW$ & Inflows $SI$, self-discharge $SD$\\
Batteries & Charging limits $PC$, self-discharge $SD$ & Inflows $SI$, waste outflows $SW$\\
\hline 
\end{tabular}
\end{center}
\end{table*}

The outcome of this optimization problem is the preliminary bids of the units on the day-ahead market, represented with the variables $P_{g,t}$, $PD_{s, t}$, and $PC_{s,t}$. For storage units with charging capabilities, the LTPO module maximizes their profit from arbitrage, scheduling the optimal charging and discharging cycles. The bids for renewable units are not optimized and follow their forecasts.

\subsection{Forming the bids}
\label{formingbids}
The outcome of the day-ahead market is driven by the merit order curve. Therefore, there is no guarantee that the bids will be cleared, honoring the $P^{min}$, or charge/discharge cycles of the units. This uncertainty has grown in recent years, as peak renewable generation often leads to low or negative prices, cutting participants out of the market~\cite{Aust2020NegativeGermany}. To counteract this, large thermal producers that cannot afford interruptions often offer negative day-ahead bids. These bids are more likely to be cleared in the market, allowing the unit to continue operating without shutting down or placing intraday bids. 

The EuroEM algorithm reflects this behavior by constructing negative bids for the thermal and storage units based on their minimum operating limits $P^{min}$, as shown in Eqs.~\ref{EuroEM_connecting_var}e -~\ref{EuroEM_connecting_var}g. To this aim, the LTPO bids ($P_{g,t}$, $PD_{s, t}$, and $PC_{s,t}$) are used to construct the final bids for the day-ahead market as shown in Eqs. ~\ref{EuroEM_connecting_var}a,~\ref{EuroEM_connecting_var}c, and ~\ref{EuroEM_connecting_var}d). Fast thermal units (e.g., fossil gas, oil) are the price-setters in many European countries \cite{EPEXSPOT2021FundamentalsMarkets}; therefore, it is in their best interest to place bids, even if the price forecast is unfavorable. If their bids are not cleared on the market, these units can quickly re-adjust their strategy and bid on the intraday market. In EuroEM, these units offer all available capacity, with respect to the minimum operating limits~\ref{EuroEM_connecting_var}b. All variables with a star superscript are passed as final bids to the day-ahead market.

\begin{subequations}
\begin{align}
    & P_{g,t}^{*, slow} = P_{g,t} - P_{g,t}^{min}, \quad \forall \{g, t\}\\
    & P_{g,t}^{*, fast} = P_{g,t}^{max} - P_{g,t}^{min}, \quad \forall \{g, t\}\\
    & PD_{s, t}^{*} = PD_{s, t} - PD_{s, t}^{min},  \quad \forall \{s, t\}\\
    & PC_{s, t}^{*} = PC_{s,t} - PC_{s,t}^{min}, \quad \forall \{s, t\}\\
    & P_{g,t}^{*, neg} = P_{g,t}^{min}, \quad \forall \{g, t\}\\
    & PD_{s, t}^{* ,neg} = PD_{s, t}^{min}, \quad \forall \{s, t\}\\
    & PC_{s, t}^{*, neg} = PC_{s,t}^{min}, \quad \forall \{s, t\}
\end{align}
\label{EuroEM_connecting_var}
\end{subequations}

\subsection{Day-ahead market module}
\label{EuroEM_dam_dam}
The day-ahead market (DAM) module provides market clearing for the system, inspired by the EPEX SPOT~\cite{EPEXSPOT2021FundamentalsMarkets}, and implements social welfare maximization (Eq. \ref{EuroEM_eq_dam}a) via linear optimization. The decision variables include: the thermal generation offered as positive ($P_{g,t}$) and negative bids ($P_{g,t}^{neg}$), the charge and discharge of storage units offered as positive ($PC_{s,t}$ and $PD_{s,t}$) and negative bids ($PC_{s,t}^{neg}$ and $PD_{s,t}^{neg}$), the hourly generation of renewable units ($P_{r,t}$) and the hourly demand ($D_{d,t}$). 

$N_D$ is the set of consumers, and the demand bids are presented with $b_{d}$. $N_R$ is the total number of renewable units in the system, and their bids, representing the VOM costs, are presented as $b_r$. The negative decision variables, $P^{neg}_{g,t}$ and $PD^{neg}_{s,t}$ are offered with negative prices, while $PC^{neg}_{s,t}$ is offered with the cost of demand shedding $b_{ds}$. This pushes them to the beginning of their respective stacks and constructs bidding curves that closely match the real market. Storage units usually act as price-takers and bid based on arbitrage opportunities and opportunity costs. To ensure their position as price takers, we set the bid prices to $\sigma_{PC} = b_{ds}$ for charging and $\sigma_{PD} = 0$ for discharging. This is similar to other approaches in literature, where sell bids for storage units are often priced at 0 \euro/MWh (price-takers)~\cite{Khaloie2024Day-AheadMarkets} and buy bids are set very high. The node balance equation presented in~\ref{EuroEM_eq_dam}b ensures that the total generation and demand are balanced at all time steps $t$ and for all control zone $i \in Z$; $F^{i,j}_{f,t}$ represents the power flows from zone $i$ to zone $j$ at time $t$. $N_F$ is the set of cross-zonal connections. Eqs.~\ref{EuroEM_eq_dam}c~-~\ref{EuroEM_eq_dam}j show the thermal and storage constraints, where the decision variables are limited by the bids formulated in subsection~\ref{formingbids}. The unbiased solution of the module is ensured by the lack of operational limits $P^{min}$ and state of charge constraints for thermal and storage units, respectively. The introduction of ramp rates in Eqs. ~\ref{EuroEM_eq_dam}d~-~\ref{EuroEM_eq_dam}e is inspired by linked block options allowed in the EPEX SPOT~\cite{EPEXSPOT2021FundamentalsMarkets}.

The constraints for the renewable units are shown in Eq.~\ref{EuroEM_eq_dam}k, where $P_{r,t}$ is the hourly generation, $P_{r,t}^{max}$ is the generation forecast. In some European markets (e.g., Germany), renewable units, and thus, their bids have higher priority \cite{FederalMinistryforEconomicAffairsandClimateActionofGermany2014Act2014}. To reflect this in the module,  Eq.~\ref{EuroEM_eq_dam}k can be modified as $P_{r, t}=P_{r,t}^{max}$ so that renewable generation will always be accepted by the market. This formulation can result in negative clearing prices when the renewable generation exceeds the demand. The demand constraint is shown in Eq.~\ref{EuroEM_eq_dam}l, where the hourly demand $D_{d,t}$ is limited by the scheduled demand $D_{sch,t}$. Finally, Eq.~\ref{EuroEM_eq_dam}m describes the power flow constraints between the control zones. The flows are bi-directional and can be limited by the net traded capacity (NTC) or flow-based limits~\cite{JAO2020ExplanationStatus}. 

\begin{subequations}
\begin{align}
    & \max_{P_g, PD_s, PC_s, P_r, D_d} \sum_{t \in T} \sum_{d \in N_D} D_{d,t} \cdot b_{d} \nonumber \\
    & + \sum_{t \in T} \sum_{s \in N_S} PC_{s,t} \cdot \sigma_{PC} + \sum_{t \in T} \sum_{s \in N_S} PC^{neg}_{s,t} \cdot b_{ds} \nonumber \\
    & - \sum_{t \in T} \sum_{g \in N_G}{P_{g,t}\cdot b_g} - \sum_{t \in T} \sum_{g \in N_G}P^{neg}_{g,t}\cdot (-b_g) \nonumber \\ 
    & - \sum_{t \in T} \sum_{s \in N_S}PD_{s,t} \cdot \sigma_{PD} \nonumber \\
    & - \sum_{t \in T} \sum_{s \in N_S}PD^{neg}_{s,t} \cdot \sigma_{PD} - \sum_{t \in T} \sum_{r \in N_R}{P_{r,t}\cdot b_r} \label{objective} \\
    & \text{s.t.} \nonumber  \\
    & Nodal \: balance \nonumber  \\
    & \sum_{g \in N_G} P^i_{g,t} + \sum_{g \in N_G} P^{i,neg}_{g,t} + \sum_{s \in N_S} PD^i_{s,t}  + \sum_{s \in N_S} PD^{i,neg}_{s,t} \nonumber \\ 
    & + \sum_{r \in N_R} P^i_{r,t} - \sum_{s \in N_S} PC^i_{s,t} - \sum_{s \in N_S} PC^{i,neg}_{s,t} - \sum_{d \in N_D} D^i_{d,t} \nonumber \\ 
    & + \sum_{\substack{f \in N_F \\i \neq j}} F^{i,j}_{f,t} = 0  , \quad \forall t, \forall i,j \in Z \\
    & Unit \: and \: system \: constraints \nonumber  \\
    & 0 \leq P_{g, t} \leq P_{g,t}^{*}, \quad \forall \{g, t\}\\
    & P_{g, t+1} - P_{g, t} \leq RU_{g}, \quad  \forall \{g, t\}\\
    & P_{g, t} - P_{g, t+1} \leq RD_{g}, \quad  \forall \{g, t\}\\
    & 0 \leq PD_{s, t} \leq PD_{s, t}^{*}, \quad \forall \{s, t\}\\
    & 0 \leq PC_{s, t} \leq PC_{s, t}^{*}, \quad \forall \{s, t\}\\
    & 0 \leq P^{neg}_{g, t} \leq P_{g,t}^{*, neg}, \forall \{g, t\}\\
    & 0 \leq PD^{neg}_{s, t} \leq PD_{s, t}^{*, neg}, \quad \forall \{s, t\}\\
    & 0 \leq PC^{neg}_{s, t} \leq PC_{s, t}^{*, neg}, \quad \forall \{s, t\}\\
    & 0 \leq P_{r, t} \leq P_{r,t}^{max}, \quad \forall \{r, t\}\\
    & 0 \leq D_{d, t} \leq D_{sch,t} \quad \forall \{d, t\}\\
    & 0 \leq F_{f, t}^{i,j} \leq F_{f}^{i,j,max} \quad \forall \{f, t\}
\end{align}
\label{EuroEM_eq_dam}
\end{subequations}

The inclusion of negative bids enables the units to fulfill their commitments and maintain their minimum operating limits during normal market operation. Unlike common market models, where negative prices can occur during grid congestion and excess renewable generation, the DAM module anticipates negative bids, resulting in behavior that better reflects the current state of the EPEX SPOT.

\section{Case study}
\label{chapter6_case_study}
\subsection{Test system}
\label{chapter6_test_system}
We benchmark the performance of the day-ahead market (DAM) dispatch with the literature standards: economic dispatch and unit commitment (presented in the Supplementary material). We employ a modified IEEE 118-bus system~\cite{IllinoisInstituteofTechnologyIIT2024IndexData} with an optimization horizon of 1 year (8,760 hours). The system contains three control zones, 118 buses, 186 branches, and nine transformers. The interconnections are between control zones 1-2 and control zones 2-3. The original system contains 54 thermal generators presented in~\cite{Pena2018AnPenetration}. In our modified version, the system contains 162 generators with the following composition:

\begin{itemize}
    \item Fast thermal units (gas, oil) - 21 units with a total installed capacity of 1,110MW.
    \item Slow thermal units (nuclear, coal, lignite) - 33 units with a total installed capacity of 4,927.5MW.
    \item Storage units (hydro dam, pumped hydro, batteries) - 54 units with a total installed capacity of 503.1MW.
    \item Renewable units (solar, wind, run-of-river hydro) - 54 units with total installed capacity of 1,006.25MW.
\end{itemize}

The total generation in the system is 7,546.85MW, while the average demand is 3,733.07MW. We assume the fast thermal and storage units can fully ramp up their generation/consumption within one hour. Additionally, $P^{min}=0$ for the fast units, as they are price-setters that often bid aggressively and provide flexibility to the market. In practice, if their bids are not accepted, they can quickly re-adjust their strategy and bid on the intraday market. The ramp rates for the slow units range from 4 hours (for hard coal and lignite) to 8 hours (for nuclear).

To ensure a reliable assessment of renewable forecasts and demand, we utilize data from France, Germany, and Switzerland to describe the three control zones in the system. We obtain the forecast for solar units, wind units, and hydro inflows from PyPSA Europe~\cite{Horsch2018PyPSA-Eur:System}. The demand profile and wholesale electricity price forecast are derived from the ENTSO-E transparency platform~\cite{ENTSO-E2022ENTSO-EPlatform} for 2021. We focus on 2021 to avoid biases caused by the Ukraine war. We provide dynamic pricing for the thermal units using their efficiency, fuel-to-energy conversion ratios, and $CO_2$ emissions per MW. We base the fuel and $CO_2$ prices on the following indices: 
\begin{itemize}
    \item Gas prices - European TTF gas index.
    \item Oil prices - since there is no single unified market for furnace oil/heavy fuel oil in Europe, we assume they are 60\% more expensive than gas, following the same price curve.
    \item Hard coal prices - Argus/McCloskey API2 index for European coal.
    \item Lignite prices - since there is no single unified market for lignite, we assume that lignite prices are 80\% cheaper than coal, following the same price curve.
    \item Uranium prices - world uranium price index URAXP.
    \item $CO_2$ emission costs - European carbon permits EU ETS index.
\end{itemize}

The bids on the day-ahead market depend on the type of unit. We assume that slow thermal units and hydro dam units must comply with the REMIT regulation by the EU (Regulation No. 1227/2011)~\cite{EuropeanUnion2011RegulationRelevance}, which prevents them from withholding energy. Therefore, they always bid truthfully based on the electricity price forecast and demand forecasts. Fast thermal units will always bid their full capacity to try and benefit from market volatility. Storage units with charging capabilities depend on market arbitrage, and in practice, they utilize more aggressive trading algorithms. To capture this in the model, we assume these units use competitive price forecasts compiled from previous market-clearing runs of the test system. We assume that all units have an uncertainty of 2.5\% for electricity and fuel prices, added as Gaussian noise. This uncertainty is reasonable and conservative, considering that day-ahead forecasting error can range up to 7.4\% as described in~\cite{Mandal2013AForecasting}. We assume that the marginal costs for renewable units are zero and that renewable curtailment incurs no costs during redispatch actions. 

Lastly, we calculate the trade capacity between control zones as the sum of the rated capacity of all interconnector lines between the zones. We assume that 40\% of this rating can be used for each direction of the interconnection to account for a 20\% transmission reliability margin.

\subsection{Security assessment setup}
\label{chapter6_test_system}
We use the AC OPF dispatch as a base scenario for the security comparison. The ancillary service market dispatch is not represented; therefore, the system control reserves are not predetermined. Instead, the available reserves are estimated for each dispatchable unit as the difference between the current dispatch and its operating limits. The redispatch model ensures that no branch violations occur based on the provided clearing; however, it does not consider the N-1 security. The comparison between the models is fair, as the OPF-based dispatch also does not account for N-1 security. We assess the system security using a list of 1,000 contingencies for 18 representative hours selected from the yearly load curve. In total, this results in 18,000 simulations of cascading failures. The list of contingencies contains single-branch failures (91\%), double-branch failures (8.3\%), and triple-branch failures (0.5\%), while the remaining contingencies involve four or more branches. To maintain parity, we use the same representative hours and list of contingencies for each test.

The EuroEM framework is developed in Python and is soft-linked with the Cascades model, which is developed in MATLAB. The simulations are executed on a workstation equipped with an AMD Ryzen™ Threadripper™ 3960X processor, with the computation times presented in Table \ref{EuroEM_execution_times}.

\begin{table}[!t]
\caption{Execution time of the different models in this study.}
\label{EuroEM_execution_times}
\centering
\begin{tabular}{|c|c|}
    \hline
    \textbf{Model} & \textbf{Execution time [s]}\\ \hline
    DAM & ~2,230 \\
    UC  & ~1,890\\
    ED & ~1,115\\
    Redispatch (every 24h) & ~19\\
    Security simulation (OPF) & ~83\\
    Security simulation (Dispatch) & ~125\\
    \hline
\end{tabular}
\end{table}

\section{Results}
\label{sec:results}
In this section, we benchmark the Day-ahead Market Dispatch model (DAM) against economic dispatch (ED) and unit commitment (UC). Finally, we examine the impact of the dispatch models on system security, comparing them to the optimal power flow (OPF), which is most often used in cascading failure analyses.

\subsection{Day-ahead market clearing}
\label{subsec:performance_comparison}
Table~\ref{EuroEM_generation_table} shows the contribution of each unit type to the total generation. We observe that the generation contribution of renewable and storage units is similar across the models. Nonetheless, notable differences are observed for some of the generation units: the ED overly relies on hard coal, as the model does not have the ability to turn the units off. Therefore, hard coal units always produce at least the equivalent of their minimum operating limits $P^{min}$. Another notable difference is the higher utilization of gas units in the DAM. Gas units take on the role of price-setters and provide flexibility to the system. Although their total contribution is less than 1\% of the total generation, gas units are activated in 5.75\% of the optimization hours in the DAM model. In comparison, gas units are activated 2.97\% of the time in the ED model and 4.54\% of the time in the UC model. Therefore, the activation of these units significantly increases the clearing price during high-demand hours, leading to more frequent price spikes in the DAM model. 

Table \ref{EuroEM_table_charging} shows how storage units are charged, underscoring the competitive dynamics in the DAM model. In ED and UC, storage units charge only to minimize the system costs. This leads to underutilization of the units, with hydro pumps barely charging over the time horizon. By contrast, charging rises across all unit types in the DAM dispatch, reflecting their efforts to exploit arbitrage opportunities. This notably improves the system's trade dynamics, leading to behaviors that better reflect the current state of European power markets.

\begin{table}
\begin{center}
\centering    
\setlength{\tabcolsep}{2pt}
\caption{Percentage of the total system generation per unit type.}
\label{EuroEM_generation_table}
\begin{tabular}{| c | c | c | c |}
\hline
Unit type & ED {[}\% of total{]} & UC {[}\% of total{]} & DAM {[}\% of total{]} \\ \hline
Oil & 0.00\% & 0.00\% & 0.00\% \\
Gas & 0.06\% & 0.08\% & 0.13\% \\
Hard Coal & 10.42\% & 6.63\% & 9.82\% \\
Lignite & 55.85\% & 59.61\% & 56.35\% \\
Nuclear & 26.84\% & 26.84\% & 26.82\% \\
Hydro dam & 2.90\% & 2.90\% & 2.90\% \\
Hydro pump & 0.27\% & 0.27\% & 0.31\% \\
Hydro pump daily & 0.00\% & 0.00\% & 0.01\% \\
Battery & 0.03\% & 0.04\% & 0.04\% \\
Run\_of\_river & 1.76\% & 1.76\% & 1.76\% \\
Solar & 1.13\% & 1.13\% & 1.13\% \\
Wind & 0.73\% & 0.73\% & 0.73\% \\
\hline 
\end{tabular}
\end{center}
\end{table}

\begin{table}
\begin{center}
\caption{Total charging of storage units in MW.}
\label{EuroEM_table_charging}
\begin{tabular}{| c | c | c | c |}
\hline
Charging technology & ED {[}MW{]} & UC {[}MW{]} & DAM {[}MW{]} \\ \hline
Hydro pump & 177 & 36 & 24,322 \\
Hydro pump daily & 1,006 & 1,009 & 5,582 \\
Battery & 10,914 & 14,980 & 17,462 \\ \hline
Total & 12,097 & 16,024 & 47,366 \\
\hline 
\end{tabular}
\end{center}
\end{table}

Table~\ref{EuroEM_table_profit} presents the cumulative costs and net profits for the ED and DAM models. The market-clearing price is extracted as the dual variable of the node balance equation. Extracting the dual variables is impossible for the UC model due to the inclusion of binary variables. In the DAM and ED models, lignite and hard coal units contribute to the majority of the costs, followed by the costs for nuclear energy. Hard coal units cannot be shut down in the ED; therefore, these units incur high costs and negative profits. In the DAM, gas units are more frequently activated, incurring higher costs and profits. This results in overall higher clearing prices for the DAM model, increasing the net profit of most units. Due to the competitive behavior of the storage units in the DAM model as described in Section~\ref{EuroEM_dam_ltpo}, they activate more often, incurring higher costs. These units also take higher risks, which improves the profit of batteries and hydro dam units, while decreasing the profit of pumped units. Hydro pumps with daily operation even incur losses due to missed trading windows. The results demonstrate that the DAM model can effectively represent a real market-clearing process by considering the bidding strategies employed by the units.

\begin{table}
\begin{center}
\centering    
\setlength{\tabcolsep}{2pt}
\caption{Total costs and net profits for the ED and DAM models.}
\label{EuroEM_table_profit}
\begin{tabular}{| c | c | c | c | c |}
\hline
Unit type & \begin{tabular}[c]{@{}l@{}}Cost ED \\ (Thsd \euro)\end{tabular} & \begin{tabular}[c]{@{}l@{}}Net profit ED \\ (Thsd \euro)\end{tabular} & \begin{tabular}[c]{@{}l@{}}Cost DAM \\ (Thsd \euro)\end{tabular} & \begin{tabular}[c]{@{}l@{}}Net profit DAM \\ (Thsd \euro)\end{tabular} \\ \hline
Oil & 0 & 0 & 0 & 0 \\
Gas & 1’498 & 64 & 3’758 & 151 \\
Hard Coal & 206’439 & -5’424 & 197’242 & 7’759 \\
Lignite & 954’406 & 96’441 & 964’754 & 128’870 \\
Nuclear & 57’137 & 443’626 & 57’137 & 457’313 \\
Hydro dam & 1’525 & 56’201 & 1’525 & 56’365 \\
Hydro pump & 141 & 4’957 & 198 & 4’394 \\
Hydro pump daily & 2 & 31 & 13 & -75 \\
Battery & 29 & 170 & 47 & 228 \\
Run\_of\_river & 0 & 30’446 & 0 & 31’124 \\
Solar & 0 & 21’193 & 0 & 21’467 \\
Wind & 0 & 13’406 & 0 & 13’602 \\ \hline
Total & 1’221’177 & 661’111 & 1’224’674 & 721’198 \\
\hline 
\end{tabular}
\end{center}
\end{table}

\begin{figure}[!t]
    \centering
    \includegraphics[width=0.5\textwidth]{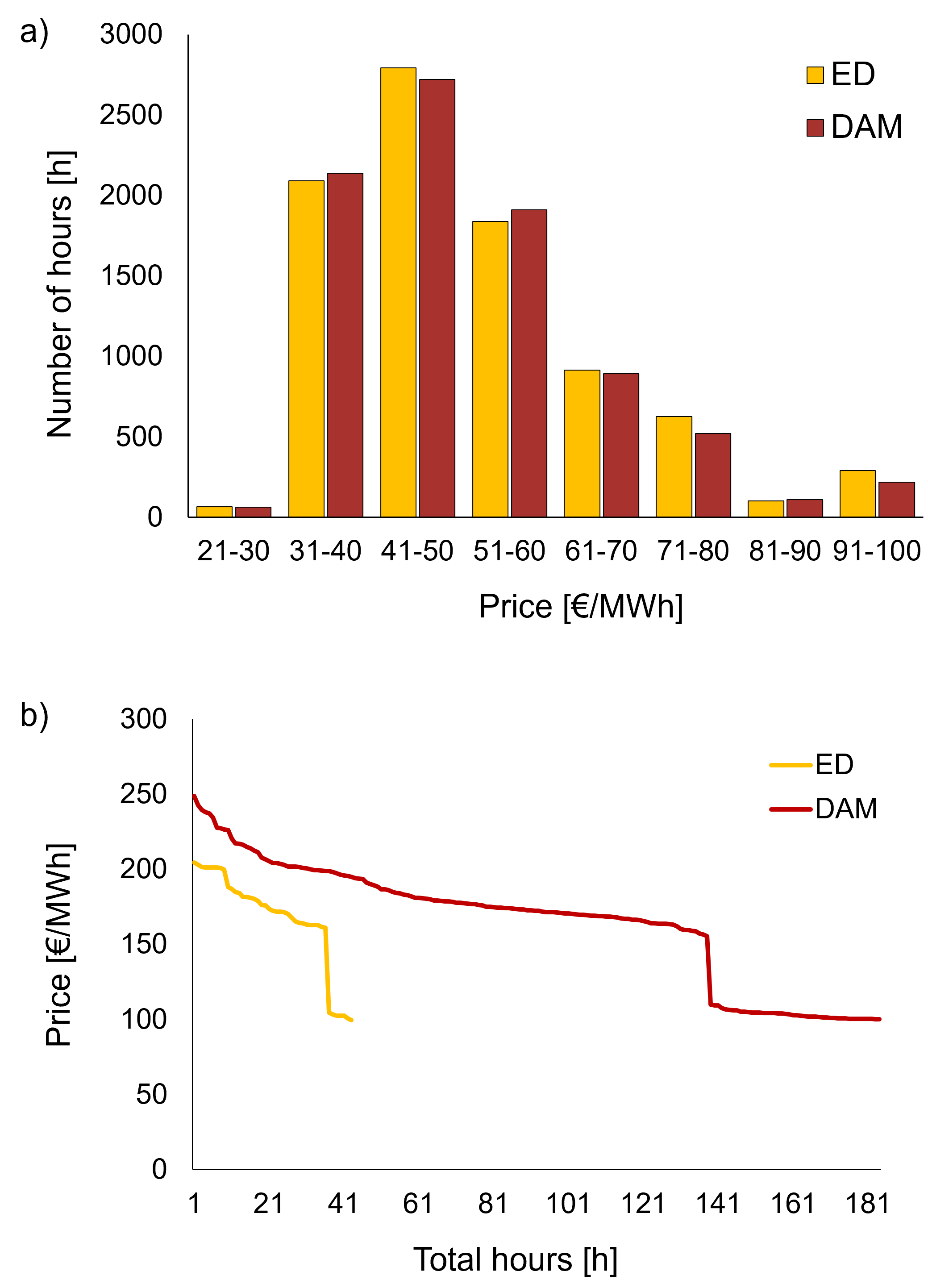}
    \vspace{-0.3cm}
    \caption{Comparison of the clearing prices in the ED and DAM models. Hours with prices above 100 \euro/MWh (high prices) are removed from the histograms to avoid disrupting the scale and are instead presented in the second subplot. a) histogram of the clearing prices of the ED and DAM; b) number of hours with prices above 100 \euro/MWh in the ED and DAM models.}
    \label{fig:EuroEM_ED_dam_price}
\end{figure}

The average clearing price difference between the two models is quite minor. The average hourly price for the ED is 51.2 \euro/MWh, while the average price for the DAM is 52.6 \euro/MWh (average increase of 2.7\%). Figure~\ref{fig:EuroEM_ED_dam_price}a shows the distribution of prices across both models, limited to prices of 100 \euro/MWh. We observe that the distribution is similar for both models, with a slight increase in the hours with prices between 91 and 100 \euro/MWh in the ED model. However, upon further inspection of the hours with prices above 100 \euro/MWh, which are generally considered high, we observe several notable differences(Figure~\ref{fig:EuroEM_ED_dam_price}b). In the DAM model, 184 hours experienced prices above 100 \euro/MWh, compared to only 42 hours in the case of ED (an increase of 4.3 times). Additionally, the highest recorded clearing price was 248.8 \euro/MWh in the DAM model, compared to 204.2 \euro/MWh in the ED model (an increase of 21.9\%). We attribute this to the more frequent activation of gas units in the DAM model, as they are often the price setters. 

Table~\ref{EuroEM_table_flows} presents the total trade capacity between the zones and the average utilized capacity for each of the models. The capacity between the zones is similarly utilized by the three models, with the lowest utilization observed by the DAM, especially in the direction of Zone 1 - Zone 2. The capacity is sufficient and does not cause price inequality between the control zones. We attribute this behavior in the DAM model to sufficient domestic production across all control zones and to the more frequent activation of storage units, which provide domestic flexibility. The Zone 2 - Zone 1 interconnection is heavily utilized by all three models, due to the large installation of cheap thermal generation in Zone 2.

\begin{table}
\begin{center}
\centering    
\setlength{\tabcolsep}{6pt}
\caption{Overview of the interconnections in the system and the average utilized trade capacity.}
\label{EuroEM_table_flows}
\begin{tabular}{| c | c | c | c | c |}
\hline
Interconnections & Capacity {[}MW{]} & ED & UC & DAM \\ \hline
Zone 1 - Zone 2 & 750 & 38\% & 33\% & 17\% \\
Zone 2 - Zone 1 & 750 & 61\% & 51\% & 41\% \\
Zone 2 - Zone 3 & 360 & 18\% & 21\% & 12\% \\
Zone 3 - Zone 2 & 360 & 29\% & 25\% & 23\% \\
\hline 
\end{tabular}
\end{center}
\end{table}

\subsection{Cost of redispatch}
The costs of redispatch are presented in Table~\ref{EuroEM_table_redispatch}. DAM results in ~15\% lower costs compared to the ED and UC models, a reduction of up to \euro326'124. The daily costs range from \euro204 to \euro12'456 for the DAM model, with an average of \euro4'909. The maximum costs incurred by ED and UC in days with very high demand are notably higher, up by a threefold. On days with low volatility, the costs for ED and UC are lower than those for DAM, as indicated by the minimum costs. The primary reason for this behavior is the frequent activation of storage and gas units in the DAM. This leads to higher domestic generation and lower trade between control zones, reducing the need for redispatch. Although the results highlight how free market behavior can reduce redispatch costs, it is important to note that we only perform N-secure redispatch, and not N-1 redispatch, which can result in unfavorable conditions for system security, as highlighted in the following subsection.

\begin{table}[]
\centering
\caption{Comparison of the redispatch costs for the dispatch models.}
\label{EuroEM_table_redispatch}
\resizebox{\columnwidth}{!}{%
\begin{tabular}{|c|c|c|c|c|c|}
\hline
\multicolumn{1}{|c|}{Model} & \multicolumn{1}{c|}{\begin{tabular}[c]{@{}c@{}}Total \\ costs (€)\end{tabular}} & \multicolumn{1}{c|}{\begin{tabular}[c]{@{}c@{}}Daily average \\ costs (€)\end{tabular}} & \multicolumn{1}{c|}{\begin{tabular}[c]{@{}c@{}}Daily min \\ costs  (€)\end{tabular}} & \multicolumn{1}{c|}{\begin{tabular}[c]{@{}c@{}}Daily max \\ costs  (€)\end{tabular}} & \multicolumn{1}{c|}{\begin{tabular}[c]{@{}c@{}}Daily 95\% \\ costs(€)\end{tabular}} \\ \hline
ED & 2’117’806 & 5’802 & 86 & 36’026 & 22’438 \\
UC & 2’006’343 & 5’497 & 88 & 36’583 & 20’642 \\
DAM & 1’791’682 & 4’909 & 204 & 12’456 & 11’159 \\ \hline
\end{tabular}%
}
\end{table}

The results highlight the validity of the proposed day-ahead market dispatch model and its usefulness in capturing the behavior of competitive electricity markets. Compared to existing methods (ED and UC), the DAM solution maintains the profit-driven nature of market participants.

\subsection{Security assessment}
\begin{figure}[!t]
    \centering
    \includegraphics[width=0.50\textwidth]{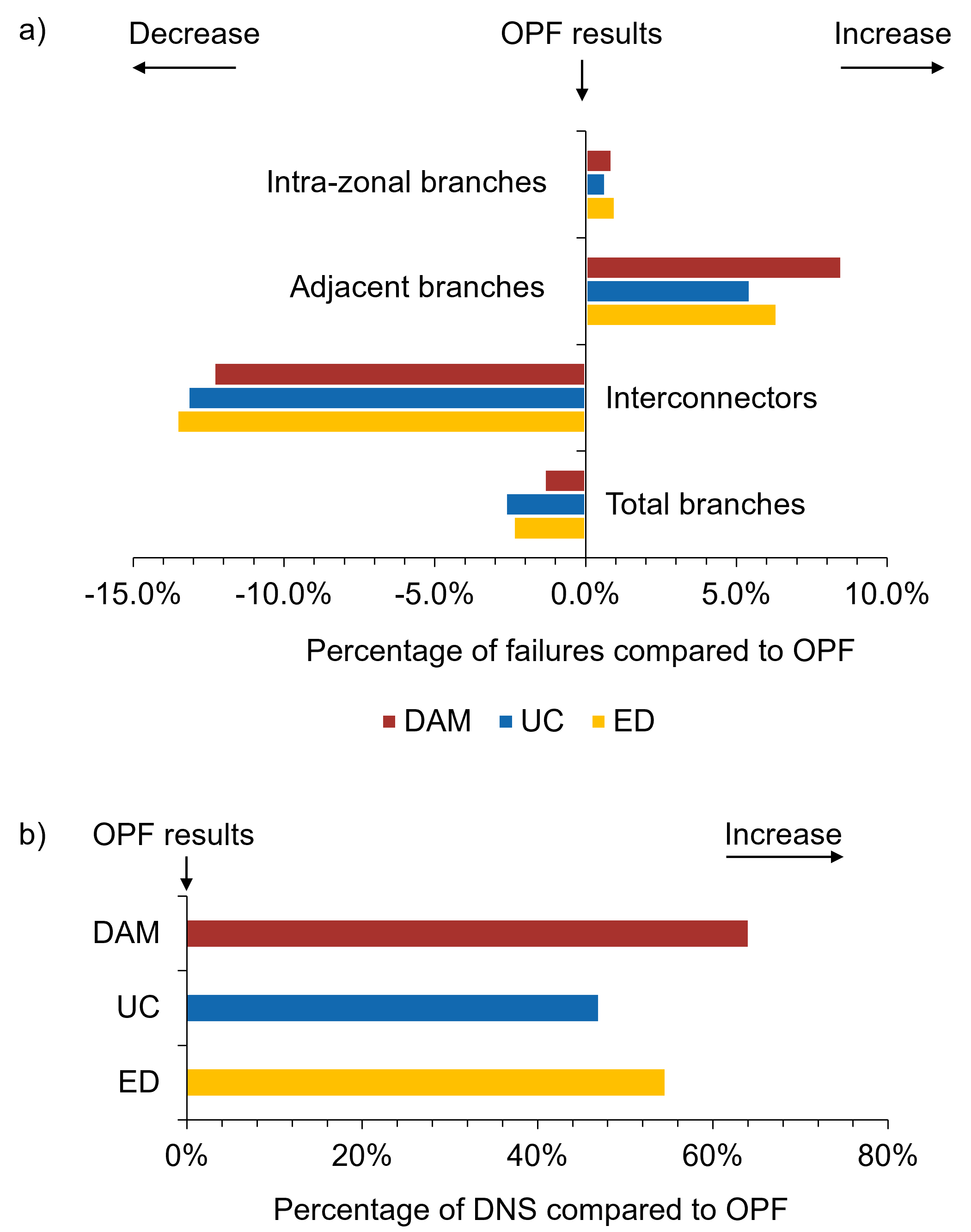}
    \vspace{-0.5cm}
    \caption{Comparison of the system performance between the DAM, ED, and UC models, and optimal power flow (OPF) based dispatch. The comparison includes a) the cumulative number of failures of total branches, interconnectors, adjacent branches, and intra-zonal branches; b) the demand not served (DNS).}
    \label{fig:EuroEM_security_bar}
\end{figure}

Figure~\ref{fig:EuroEM_security_bar}a shows a comparison of the branch failures between the three dispatch models and OPF-based dispatch. All dispatch methods result in a lower number of total branch failures, primarily driven by a reduction in interconnector failures. Despite this, the dispatch methods result in an increased number of adjacent and intra-zonal branches. Adjacent branches are the intra-zonal branches connected to an interconnector. These branches are the recipients of cross-zonal flows and are often weak transmission links in the system~\cite{Stankovski2022Multi-zonalSystems}. The increase in intra-zonal branch failures also indicates larger-scale cascading progression within the control zones. The largest number of adjacent branch failures is observed with the DAM model, indicating an unfavorable generation distribution post-redispatch across the control zones, which forces the system to operate near critical conditions.

Ultimately, this significantly affects the cumulative demand not served (DNS) in the system, as highlighted in Figure~\ref{fig:EuroEM_security_bar}b. The DNS drastically increased for all dispatch models, with the highest increase observed with the DAM model. The DAM model produced a cumulative DNS that was 64\% higher than in OPF. By comparison, the increases were 54\% in ED and 47\% in UC. The high utilization of storage and gas units in the DAM provides generation flexibility in the system, but also increases the utilization of intra-zonal branches. In the event of large-scale disruptions, this places the system in critical loading conditions, supporting the development of cascading failures.

\begin{figure}[!t]
    \centering
    \includegraphics[width=0.45\textwidth]{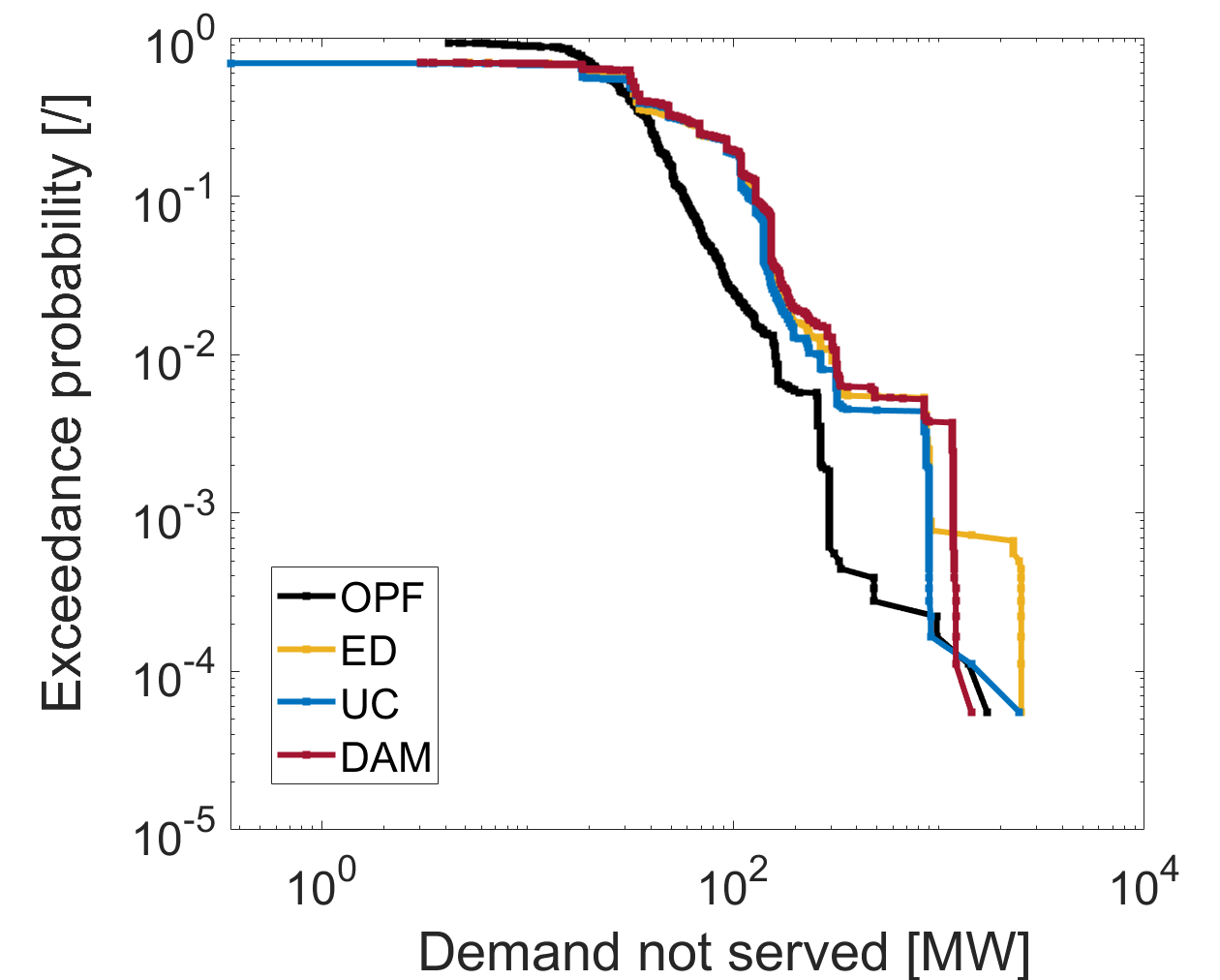}
    \vspace{-0.2cm}
    \caption{Risk curves of the cascading simulation model using OPF (black), economic dispatch (ED, yellow), unit commitment (UC, blue), and day-ahead market dispatch (DAM, red). The curves that are more pronounced on the right side exhibit worse cumulative demand not served DNS. The models use the same set of cascade-initiating contingencies in the four assessments.}
    \label{fig:EuroEM_risk_curve}
\end{figure}

The risk curves of the cascading failure analysis (Figure~\ref{fig:EuroEM_risk_curve}) or complementary cumulative distribution functions quantify the probability of observing an event with a larger DNS than the value on the x-axis. We use the criteria outlined in ~\cite{Stankovski2023PowerEvidence} to classify the size of the events. The dispatch models have lower DNS than OPF only during minor events, with DNS of 25MW or lower. These events have a very high probability and minor DNS impact, as observed on the left-hand side. This supports the findings from Figure~\ref{fig:EuroEM_security_bar}a, in which OPF had the highest number of branch failures. As most failure events observed in the system are minor, the number of branch failures is highest with OPF. However, the dispatch methods significantly compromise system security during low-probability, high-impact events (DNS above 100MW). This is especially true for the day-ahead market model, as it results in the highest cumulative DNS. Surprisingly, the ED and UC models yielded the same worst-case cascading event, as indicated by the tail of the curve. This event occurred when the system operated near critical conditions, with a total demand of 3'546 MW. The resulting cascade led to six consecutive failures and a 71\% loss of the total system demand. In contrast, this cascading event did not cause the same damage in the DAM dispatch, resulting in a minor loss of system demand. The worst event in the DAM dispatch occurred during an hour with a demand of 5'574 MW, when two simultaneous failures led to the shedding of 26\% of the total demand.

These findings are significant, as OPF solutions result in more stable systems by accounting for grid constraints when determining the optimal dispatch. This, however, does not accurately represent the operation of many real power systems, where markets are decoupled from grid operations. Therefore, system security is significantly overestimated when using OPF-based dispatch.

\section{Conclusions}
\label{sec:conclusions}

In this paper, we introduce a day-ahead market (DAM) dispatch model capable of simulating the profit-driven nature of the European power market. As part of the EuroEM framework, the model optimizes long-term positions to identify the most profitable bids for market participants. The bids are then cleared in a merit-order process, resulting in a true market dispatch. We introduce methods to ensure that all unit constraints are satisfied without compromising market independence. To simulate the actions of the TSO, we simulate the market dispatch to ensure it respects the grid constraints. Lastly, we test the model on a modified IEEE-118 bus system, where we analyze the impact of market dispatch on system security.

Our results show that the DAM model provides a dispatch comparable to that of the ED and UC models, with several notable differences. Dispatch of slow thermal units, renewables, and hydro dam units is almost identical across the models. However, the DAM led to notably higher utilization of gas units and storage units with charging capabilities. This results in a clearing price that is 2.7\% higher on average, and a 4.3 times increase in the frequency of hours with very high prices. The higher prices improve the economic performance of thermal units; the performance of renewable and hydro dam units is identical to that of the ED model, observing minor economic gains. Batteries also mark significant improvements in net profits, attributed to the more frequent arbitrage activations. We also observe how missed trading windows can result in economic losses, such in the case of hydro pumps with daily operation. Finally, we observe that the DAM dispatch results in lower redispatch costs compared to ED and UC due to the generation provided by gas and storage units.
The security analysis shows that system security is severely overestimated when using a standard OPF-based dispatch. The market-informed dispatch results in a significant increase in adjacent branch failures. Additionally, we observe a 64\% increase in the cumulative DNS in the case of the DAM model (Figure~\ref{fig:EuroEM_security_bar}b). This is especially true of high-impact cascading events that severely undermine grid security. Despite efforts to avoid congestion, the decoupled market approach can force the system to operate near critical conditions due to an unfavorable generation distribution, resulting in increased loading on adjacent lines. As a result, the proposed DAM model offers an in-depth understanding of the impacts of market clearing on system security.

In our future work, we will expand the EuroEM framework by introducing a futures and balancing (ancillary services) market. Additionally, we aim to further improve the day-ahead market bidding process by introducing a portfolio optimization model and transitioning to a 15-minute resolution. The purpose of the model is to capture the behavior of large asset owners that dominate the real electricity market. Finally, we intend to validate the performance of the EuroEM framework against the real EPEX SPOT by utilizing large datasets of European generation assets.

\section*{Acknowledgements}
\label{Acknowledgements}
The research published in this publication was carried out with the support of the Swiss Federal Office of Energy as part of the SWEET consortiums EDGE and PATHFNDR. The authors bear sole responsibility for the conclusions and the results presented in this publication. We would like to acknowledge Maxime Descamps for his help with data processing and model development.

\section*{Data and code availability}
\label{Acknowledgements}
We encourage modeling and data transparency. The model, data, and additional processing tools used in this work are freely available and can be accessed in the following public repository: \url{https://gitlab.ethz.ch/andrejst/rre-power-market-model}. Further information and requests for resources and materials should be directed to and will be fulfilled by the lead contact, Giovanni Sansavini (sansavig@ethz.ch).

\bibliographystyle{IEEEtran}
\bibliography{refs}

\end{document}